\documentclass[prc,showpacs,showkeys,superscriptaddress,twocolumn,floatfix,nofootinbib]{revtex4}
\pdfoutput=1
\usepackage[pdftex]{graphicx}
\usepackage{amssymb,amsmath,bm,color}
\usepackage[sort&compress]{natbib}
\def\vec#1{{\bf #1}}
\def\Im{{\rm Im\;}}\def\Re{{\rm Re\;}}
\begin{document}

\title{Are back-to-back  particle--antiparticle correlations observable
  \mbox{in high energy nuclear collisions?}}

\author{J\"o{}rn
  Knoll}\thanks{e-mail:j.knoll@gsi.de} \affiliation{GSI
  Helmholtzzentrum f\"ur Schwerionenforschung GmbH, Planckstr. 1, 64291
  Darmstadt, Germany}

\begin{abstract}
  Analytical  formulae   are  presented  which   provide  quantitative
  estimates  for  the  suppression  of  the  anticipated  back-to-back
  particle--antiparticle   correlations   in   high   energy   nuclear
  collisions due  to the finite  duration of the  transition dynamics.
  They show that it is unlikely to observ the effect.
\end{abstract}

\date{\today}
\maketitle

\section{Introduction}
In  1996 Asakawa  and  Cs\"o{}rg\H{o} \cite{Asakawa:1996xx}  suggested
that   back-to-back  particle--antiparticle  correlations   should  be
observable  in  high-energy   nuclear  collisions,  see  also  earlier
considerations in  Refs.  \cite{Vourdas:1988cd,Razumov:1994xu}.  Based
on a sudden transition assumption huge effects were predicted for this
phenomenon.   About  a  further  dozen applications  followed.   In  a
subsequent paper  together with Gyulassy  \cite{Asakawa:1998cx} it was
shown that  the finite duration  $\tau$ of the transition  reduces the
effect.  In that paper, however,  the authors used a discontinuous and
therefore unrealistic transition profile that led to large ultraviolet
Fourier components and thus to a very moderate reduction of the effect
of order $(2\omega\tau)^{-2}$.  Here  $2\omega$ is the energy required
to produce the pair.  In this short  note it will be shown that with a
smooth  transition  profile  and  realistic transition  durations  the
suppression is  rather exponential, $\propto  {\rm e}^{-4\omega\tau}$,
deferring to observe the effect.
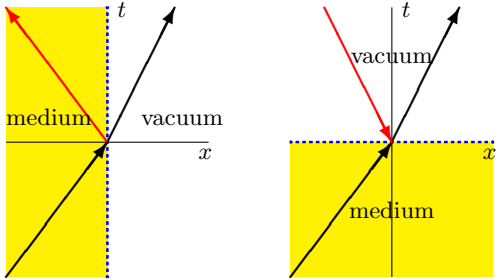
\begin{figure}[b]
\unitlength0.0025\textwidth
\hfill
\begin{picture}(60,85)
\put(0,5){
\put(15,40){\makebox(0,0){\color{yellow}\rule{30\unitlength}{80\unitlength}}}
\thicklines
\put(30,0){\color{blue}\dashbox{1}(0,80)}
\put(0,0){\vector(3,4){30}}
\put(30,40){\vector(1,2){20}}
\put(30,40){\color{red}\vector(-3,4){30}}
\thinlines
\put(0,40){\line(1,0){60}}
\put(57,35){$x$}
\put(33,77){$t$}
\put(0,45){medium}
\put(40,45){vacuum}
}
\end{picture}
\hfill
\begin{picture}(60,80)
\put(0,5){
\put(30,20){\makebox(0,0){\color{yellow}\rule{60\unitlength}{40\unitlength}}}
\put(30,0){\line(0,1){80}}
\thicklines
\put(0,40){\color{blue}\dashbox{1}(60,0.1)}
\put(0,0){\vector(3,4){30}}
\put(30,40){\vector(1,2){20}}
\put(10,80){\color{red}\vector(1,-2){20}}
\put(57,35){$x$}
\put(33,77){$t$}
\put(30,65){\makebox(0,0){vacuum}}
\put(30,20){\makebox(0,0){medium}}
}
\end{picture}
\hfill\hspace*{1mm}
\caption{Optical  ray  pictures  in  space-time.  Left:  the  standard
  reflection-transmission situation,  where a wave  traverses from one
  medium  (yellow)  to  another   one  (e.g.   vacuum)  at  a  spatial
  interface.  Right: the sudden transition  in time from the medium to
  the vacuum; here the frequency $\omega$ is discontinuous, creating a
  second component with negative $\omega$ which has the interpretation
  of an  antiparticle with opposite momentum.  The  (blue) dashed line
  separates the media, the  arrow sense distinguishes between particle
  and antiparticle.   By means of Lorentz  transformations the concept
  can  directly be  generalized  to hyper-planes  and  even to  curved
  hyper-surfaces.}\label{optics}
\end{figure} 

\section{The sudden picture}

The main assumption of  the original approach \cite{Asakawa:1996xx} is
the sudden change from the in-medium situation to that in vacuum.  The
authors used  the Bogolioubov-Valatin (BV)  transformation to describe
the effect,  a picture that  may not be  so intuitive for many  of us.
Indeed    the   effect    is    nothing   else    than   a    standard
``reflection--transmission'' problem at the  interface of two media, a
well  known  problem  in  physics,  in particular  in  optics.   Thus,
transcribing  our  wisdom from  the  spatial  situation,  like a  wave
traversing from  one medium to another  one at a  sharp interface (see
Fig.~\ref{optics} left)  to the  here considered sudden  transition in
time  (Fig.~\ref{optics}   right),  precisely  recovers   the  results
presented in \cite{Asakawa:1996xx,Asakawa:1998cx}. The common boundary
condition  is that  one has  one  incoming wave  and two  time-forward
propagating    outgoing     wave    components.     As     in    Refs.
\cite{Asakawa:1996xx,Asakawa:1998cx}  we  restrict  the discussion  to
(charged)  relativistic  bosons  described  by the  Klein-Gordon  (KG)
equation.  For  the sudden case  the wave functions with  positive and
negative energies prior and post  to the sudden transition can then be
written as (using units with $\hbar$=c=1)
\def\i{{\rm i}}
\begin{eqnarray}
\Psi_{\rm med}^{\pm}(x)&=&\frac1{\sqrt{2\Omega_{\vec{k}}}}
{\rm e}^{\mp \i \Omega_{\vec{k}} t +\i\vec{k}\vec{x}}\quad{\rm
  for}\quad t<0
\ ,\\
\Psi_{\rm vac}^{\pm}(x)&=&\frac1{\sqrt{2\omega_{\vec{k}}}}\left(
c_{\vec{k}}{\rm e}^{\mp \i \omega_{\vec{k}} t +\i\vec{k}\vec{x}}
+s_{\vec{k}}{\rm e}^{\pm \i \omega_{\vec{k}} t +\i\vec{k}\vec{x}}\right).
\end{eqnarray}
Here $\Omega_{\vec{k}}$ and $\omega_{\vec{k}}$ are the single-particle
energies  at  momentum $\vec{k}$  in  the  two  media.  Continuity  of
$\Psi(x)$ and of $\partial\Psi(i)/\partial t$ at transition time $t=0$
determines the coefficients on the vacuum side to
\begin{eqnarray}
c_{\vec{k}}&=\frac12
\left(\sqrt{\frac{\omega_{\vec{k}}}{\Omega_{\vec{k}}}}
+\sqrt{\frac{\Omega_{\vec{k}}}{\omega_{\vec{k}}}}\right)
=&\cosh[r_{\vec{k}}]\ ,\\
s_{\vec{k}}&=\frac12
\left(\sqrt{\frac{\omega_{\vec{k}}}{\Omega_{\vec{k}}}}
-\sqrt{\frac{\Omega_{\vec{k}}}{\omega_{\vec{k}}}}\right)
=&\sinh[r_{\vec{k}}]\ ,\\
&{\rm with}\quad r_{\vec{k}}
=\frac12\ln\frac{\omega_{\vec{k}}}{\Omega_{\vec{k}}},&
\end{eqnarray}
while due  to the spatial  homogeneity the spatial  momentum $\vec{k}$
remains  unchanged.  Within  the standard  relativistic  quantum field
theory  (RQFT) interpretation  of  the negative  energy components  as
antiparticles with  opposite spatial momenta,  cf.  Fig.  \ref{optics}
right frame,  this result identically  reproduces that given  in Refs.
\cite{Asakawa:1996xx,Asakawa:1998cx}   within  the  BV   picture.   In
particular the  ratio of the  antiparticle over particle  component on
the vacuum side
\begin{eqnarray}
{\cal A}=\left|\frac{s_{\vec{k}}}{c_{\vec{k}}}\right|
        =\left|\frac{\Omega_{\vec{k}}
         -\omega_{\vec{k}}}{\Omega_{\vec{k}}+\omega_{\vec{k}}}\right|
\label{sudden}
\end{eqnarray}
is the analog of the well known reflection coefficient for a wave
traversing from one medium to another (Fig. \ref{optics} left)
\begin{eqnarray}
    {\cal R} &=& \left|\frac{K-k}{K+k}\right|,
\label{reflection}
\end{eqnarray}
where $K$ and $k$ denote the  moduli of the spatial momenta in the two
media.   In the  spatial  case solely  ${k}_{\perp}$ is  discontinuous
($\omega$  and $\vec{k}_{\|}$  are continuous).   In the  sudden case,
however,  $\vec{k}$ is  continuous,  while the  discontinuity in  time
causes a discontinuity in $\omega$ thereby creating a second component
with negative  $\omega$, i.e.   an antiparticle with  opposite spatial
momentum $\vec{k}$  (Fig.  \ref{optics} right  frame).  This mechanism
led  to the  back-to-back  particle--antiparticle correlation  picture
advocated in Ref.   \cite{Asakawa:1996xx}.  Besides small antiparticle
components arising form the existing particles in the medium, the time
dependent interaction  can also create  genuine particle--antiparticle
pairs out  of the vacuum. The  latter, which is the  bosonic analog to
the Dirac  case, where a  time-dependent interaction pulse can  lift a
particle from  the filled Dirac sea to  the particle space\footnote{In
  the Dirac  case also  spatial transitions (Fig.   \ref{optics} left)
  lead  to  spon\-taneous  pair  production, known  as  Klein  paradox
  \cite{Klein:1929zz}, once the vector  potential changes by more than
  twice the rest  mass.}, can formally be included  by adding a ``+1''
to the  boson occupations  $n_{\vec{k}}$ of the  antiparticles.  After
the transition the one-body density becomes
\begin{eqnarray}
n_1(\vec{k})&=&|c_{\vec{k}}|^2n_{\vec{k}}
              +|s_{\vec{k}}|^2(n_{\vec{k}}+1)\\
n_{\vec{k}}&=&\frac{1}{\exp(|\Omega_{\vec{k}}|/T)+1},
\end{eqnarray}
while the back-to-back  correlation function of particle--antiparticle
pairs over the product of single yields becomes \cite{Asakawa:1998cx}
\begin{eqnarray}
\hspace*{-5mm}
C_2(\vec{k},-\vec{k})&=&\frac{n_2(\vec{k},-\vec{k})}
                           {n_1(\vec{k})n_1(-\vec{k})}\\
&=&1+\frac{|c_{\vec{k}}^*s_{\vec{k}}n_{\vec{k}}
         +c_{-\vec{k}}^*s_{-\vec{k}}(n_{-\vec{k}}+1)|^2}
          {n_1(\vec{k})n_1(-\vec{k})}\\
&\hspace*{-7mm}
_{\mbox{$\longrightarrow$}\atop|s_{\vec{k}}|\ll 1}\!\!\!\!&  
        1+ |s_{\vec{k}}|^2
\left|\frac{2n_{\vec{k}}+1}{n_{\vec{k}}
   +|s_{\vec{k}}|^2(n_{\vec{k}}+1)}\right|^2\!.
\label{C2sk}
\end{eqnarray}
Due  to   the  sudden   pair  creation  processes   this  back-to-back
correlation  ratio  can  attain  huge  values,  once  the  statistical
occupations   $n_{\vec{k}}$   fall   below   $|s_{\vec{k}}|^2$.    The
subsequent considerations  will show that  this effect is  an artifact
resulting from the sudden limit.

\section{The continuous transition case}

Compared     to     the     BV    transformation     considered     in
\cite{Asakawa:1996xx,Asakawa:1998cx}  the wave dynamical  picture used
here can  easily be generalized  to the continuous transition  case by
solving the time dependent Klein-Gordon equation (suppressing in the
following the dependence on  spatial momentum
$\vec{k}$)
\begin{eqnarray}
\partial_t^2 \Psi(t)+\Omega^2(t)\Psi(t)=0
\label{KG-EQ}
\end{eqnarray}
for a smoothly time-dependent  dispersion relation
\begin{eqnarray}
  \Omega^2(t)&=&\!{\omega^2
  +\Pi^{\rm R}(t)}
 =\!{\omega^2+\Pi^{\rm R}(-\infty)  F(t)}.
\label{def:F(t)}
\end{eqnarray}
Here  negative  times refer  to  the  situation  in the  medium  where
$\Pi^{\rm  R}$   denotes  the  time-dependent   retarded  polarization
function\footnote{In general the  polarization function $\Pi$ can also
  be  non-local  in   time  and  thus  be  energy   dependent  in  the
  semi-classical interpretation.  Then the  mapping from the medium to
  the   vacuum  case   will  involve   corresponding  ``$z$''-factors,
  $z={1}/(1-{\partial\Re\Pi^{\rm  R}}/{\partial  \omega^2})$, as  e.g.
  shown    in    Kadanoff-Baym    final-state   interaction    picture
  \cite{Knoll:2010}.   For the  example cases  discussed here  we will
  discard such  generalization and stick to time  local cases.}, while
the vacuum case is attained  for large positive times. Thereby $F(t)$,
cf.  Fig.  \ref{F-t}  below, will  later be  used to  parameterize the
transition profile.

Besides  direct  numerical   solutions  of  (\ref{KG-EQ}),  there  are
approximate analytical  methods to obtain  the reflection coefficient.
These  concern  the  single   reflection  limit  as  well  as  refined
semi-classical methods.

\subsection{Single reflection approximation}
A  simple generalisation  of  the sudden  limit  towards a  contineous
treatment is provided by the single reflection approximation (SR).  In
this  scheme  the continuous  function  $F(t)$  is  approximated by  a
sequence of steps each  providing a reflected wave component according
to (\ref{sudden}).  In  the continuum limit the coherent  sum of these
partial reflections leads to
\begin{eqnarray}
{\cal A}^{\mbox{\tiny SR}}&=&\int_{-\infty}^{\infty}\! dt\ 
\frac{\dot{\Omega}(t)}{2\Omega(t)}
\exp\left(-2\i\int_{0}^{t}dt'\ \Omega(t')\! \right)\\
&\approx&\int_{-\infty}^{\infty}\! dt\ 
\Psi_{\rm f}^{*}(t)\dot{\Omega}(t)\Psi_{\rm i}(t).
\end{eqnarray}
Thereby the  appropriate phase coherence is  approximately provided by
describing the unreflected wave  components of particles (i) and their
antiparticle partners (f) in the semi-classical (WKB) limit as
\begin{eqnarray}
\Psi_{\rm
  i}(t)=\Psi_{\rm f}^{*}(t)  \approx
\textstyle\exp(-i\int^t_{0} dt' \Omega(t'))/
\sqrt{2\Omega(t)}.
\label{PsiWKB}
\end{eqnarray}
Since $\dot{\Omega}$ is  peaked and limited to a  narrow range in $t$,
the antiparticle amplitude  $\cal A$ is essentially given  by the time
Fourier transformed of $\dot{\Omega}(t)$, i.e.
\begin{eqnarray}
\!\!|s_{\vec{k}}|&\approx&\left|{\cal A}^{\mbox{\tiny SR}}\right|
\approx \left|\frac{\Omega-\omega}
  {\Omega+\omega}\right||g(2\bar{\omega})|
\quad{\rm with}
\label{singleR}
\\
\!\!\!g(\omega)&=&\!\int\! dt\ e^{-\i\omega t}f(t),\quad
\ f(t)=-\frac{d F(t)}{dt}.
\end{eqnarray}
with         abbreviations         $\Omega=\Omega(-\infty)$        and
$\omega=\Omega(+\infty)$.  There\-by $\bar{\omega}=\Omega(\bar{t})$ is
chosen around the maximum  of $\dot{\Omega}$.  As multiple reflections
at  the  different  steps  are suppressed,  the  approximation  solely
recovers terms linear in $\Delta\Omega=\Omega-\omega$. Compared to the
sudden  result  (\ref{sudden})  the  suppression of  the  antiparticle
amplitude  caused by  the finite  duration of  the transition  is then
given  by $g(2\bar\omega)$.   This was  also the  approximation scheme
used  in  Ref.   \cite{Asakawa:1998cx}  where, however,  a  completely
unrealistic    form     for    $f(t)$    was     considered,    namely
$f(t)=\Theta(t)\exp[-t/\tau]/\tau$.  This  form contains a  sharp jump
which  leads  to unrealistically  high  Fourier  components for  large
$\omega$, and thus only to  a very moderate suppression of the effect!
More  realistic   forms  for   $\Omega(t)$  generally  will   lead  to
exponential   suppression  factors,   as  e.g.    for   the  following
analytically solvable cases
\begin{figure}[b]
\vspace*{-5mm}
\includegraphics[angle=0,width=0.5\textwidth]{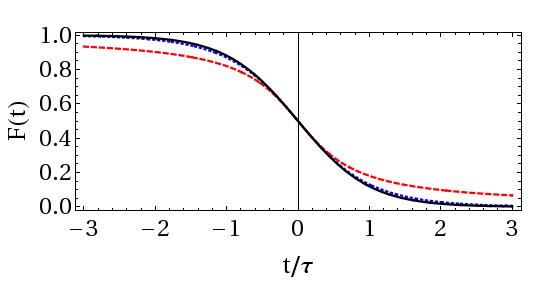}
\vspace*{-5mm}
\caption{(Color online)  $F(t)$ for the  three example cases  of table
  (\ref{table}),  (a) red  dashed,  (b) blue  dotted,  (c) black  full
  line.}\label{F-t}
\end{figure}
\begin{equation}
\begin{array}{|c|c|c|c|}
\hline
&F(t)&f(t)&g(\omega)\\ \hline
{\rm (a)\vphantom{\displaystyle\int}}&
\frac12-\frac{1}{\pi}\arctan\frac{\pi t}{2\tau}&
\frac{2\tau}{\pi^2t^2+4\tau^2}&
{\rm e}^{-|2\omega\tau/\pi|}
\\ \hline
{\rm (b)\vphantom{\displaystyle\int}}&
\frac12-\frac{1}{\pi}\arctan\!\left(\sinh\! \frac{\pi t}{2\tau}\right)&
\frac{1}{2\tau\cosh\left({\pi t}/{(2\tau)}\right)}&
\frac{1}{\cosh\left({\omega\tau}\right)}
\\ \hline
{\rm (c)\vphantom{\displaystyle\int}}&
\frac{1}{\exp(2t/\tau)+1}&
\frac{1}{2\tau\cosh^2({t}/{\tau})}&
\frac{\pi \omega\tau}{2\sinh({\pi\omega\tau}/{2})}
\\ \hline
\end{array}\label{table}
\end{equation}
In order  to assure comparable  results all functions are  chosen such
that  $F(t)$ monotonically  falls  from 1  to  0 with  a maximum  time
derivative  at   $t=0$  of  $f(0)=-\dot{F}(0)=1/(2\tau)$,   cf.   Fig.
\ref{F-t}.   Then $\tau$ is  approximately the  ``half time''  for the
change in $\Omega^2(t)$.  For short transitions times $\tau\rightarrow
0$,  one   verifies  the  sudden  result   (\ref{sudden})  since  then
$g(2\omega)\rightarrow  1$.  For  large  $\bar{\omega}\tau$ all  three
cases indeed lead to exponential suppression factors
\begin{eqnarray}
|g(2\bar\omega)|^2\ 
\raisebox{-0.6em}{$\stackrel{\textstyle\longrightarrow}
         {\scriptstyle\bar{\omega}\tau\gg 1}$ }\ 
C\ {\rm e}^{-4\alpha\bar\omega\tau}
\label{g2omega}
\end{eqnarray}
with  $C=(1,4,(4\pi\bar{\omega}\tau)^2)$  and $\alpha=(2/\pi,1,\pi/2)$
for the  cases (a)  to (c) of  (\ref{table}).  Thereby  the asymptotic
exponential behavior  is directly determined by the  imaginary part of
the nearest complex  pole position of $F(t)$.  

The physical case  where the polarization function is  switched on and
off, i.e  where $F(t)$ takes  the form of  a bell shape can  simply be
retrieved by replacing $F(t)$ by its time derivative $F_{\rm bell}(t)=
2\tau f(t)$  with $f(t)$ from table  (\ref{table}).  The corresponding
``pulse'' has  a maximum  of 1  and a time  integral of  $2\tau$.  The
suppression factor  (\ref{g2omega}) for the pair  production then will
simply     attain     an     additional     pre-exponential     factor
of $(4\bar{\omega}\tau)^2$.

\subsection{Complex semi-classical method (CWKB)}

An  alternative  method to  access  the  reflected  wave component  is
provided by the {\em complex} WKB method \cite{Knoll:1975zs}.  In that
case  the  ``reflected''  wave  component results  from  the  analytic
continuation of the ``unreflected''  WKB wave $\Psi_{\rm i}$, cf.  Eq.
(\ref{PsiWKB}), around the nearest  {\em complex valued} turning point
$t_0$ where $\Omega^2(t_0)=0$.  Then the reflection coefficients $\cal
R$ or $\cal A$ are given by \cite{Knoll:1975zs}
\begin{eqnarray}
\!\!{\cal A}^{\mbox{\tiny CWKB}}=\exp[-2|\Im S(t_0,t)|] 
\quad (\mbox{for real }t),
\label{A-CWKB}
\end{eqnarray}
where 
\begin{eqnarray}
S(t_0,t)=\int^{t}_{t_0}\!\!\! dt'\, \sqrt{\Omega^2(t')},\ 
\label{S0(t)}
\end{eqnarray}
is the action integral from the nearest complex turning point $t_0$ to
some time  $t$ on the real  axis. Since the action  integral along the
real axis does not contribute to  $\Im S$, a rough estimate can simply
be obtained  by choosing the  integration contour starting  from $t_0$
just parallel to the imaginary axis till the real axis.  Assuming that
along this  contour the integrand  can essentially be  approximated by
its value on the real axis one obtains the estimate
\begin{eqnarray}
\Im S(t_0)\approx\Omega(\Re t_0)\ \Im t_0\ .
\end{eqnarray}
For the three analytic cases  of table (\ref{table}) one verifies that
the complex turning points are  located close to the pole positions of
$F(t)$ with  values for $\Im t_0=\alpha\tau$ with  $\alpha$ = $2/\pi$,
$1$  and   $\pi/2$,  respectively  for   the  cases  (a)  to   (c)  in
(\ref{table}).  Thus the pair creation rate then becomes
\begin{eqnarray}
|s_{\vec{k}}|^2\approx|{\cal A}^{\mbox{\tiny CWKB}}|^2\approx
 e^{-4\alpha\bar{\omega}\tau}
\label{skWKB}
\end{eqnarray}
in  agreement  with  the  leading  exponential terms  of  the  single
reflection approximation given in  (\ref{g2omega}).

\subsection{The exact Fermi function case}
The  Fermi-function case  (\ref{table}c) is  particularly interesting,
since  the corresponding KG  equation (\ref{KG-EQ})  can be  solved in
closed form, providing the exact result \cite{Fluegge:65}
\begin{eqnarray}
\hspace*{-1cm}{\cal A}^2_{\rm Fermi}\!\!&=&\!\!\displaystyle
\left|\frac{\sinh(\pi(\Omega-\omega)\tau/2)}
{\sinh(\pi(\Omega+\omega)\tau/2)}\right|^2 .\label{Fermi}
\end{eqnarray}
It  generalizes the  sudden  result (\ref{sudden})  to the  continuous
Fermi-function case  (\ref{table}c) and further  confirms the limiting
cases of  the single-reflection approximation  (\ref{singleR}) and the
CWKB  result (\ref{skWKB})  within  their validity  ranges.  Also  the
corresponding  action integral  (\ref{S0(t)}) and  thus  its imaginary
part can be obtained in closed form
\begin{eqnarray}
\hspace*{-5mm}
S_{\rm Fermi}(t_0,t)=
\Omega\tau\ {\rm artanh}{\frac{\Omega(t)}{\Omega}}
-\omega\tau\ {\rm artanh}{\frac{\Omega(t)}{\omega}}\\
|\Im S_{\rm Fermi}(t_0,t)|=
\frac{\pi}{2}\; {\rm Min}(\Omega,\omega)\ \tau\quad\ (\mbox{for real }t),
\end{eqnarray}
the imaginary  part resulting from that artanh  function with argument
larger than  1.  The corresponding CWKB  amplitude (\ref{A-CWKB}) then
becomes
\begin{eqnarray}
\left|{\cal A}^{\mbox{\tiny CWKB}}_{\rm Fermi}\right|^2
&=
&\!\!\exp\left[-2\pi\; {\rm Min}(\Omega,\omega)\ \tau\right],
\label{sc-Fermi}
\end{eqnarray}
in full  compliance with the semi-classical  limit $(\omega\tau\gg 1$,
$\Delta\Omega\tau\gg 1$) of the exact Fermi-function result (\ref{Fermi}).

\section{Summary and concluding remarks}

For  nuclear  collisions  with  a  freeze-out temperature  $T$  and  a
correspondingly  smooth time dependence  of the  polarization function
$\Pi$, the back-to-back correlation (\ref{C2sk}) can then be estimated
to
\begin{eqnarray}
\left|c_2(\vec{k},-\vec{k})-1\right|&\approx& 
{\cal O}({|s_{\vec{k}}|^2}/{n_{\vec{k}}^2})
\approx{\rm e}^{-\Omega_{\vec{k}}(4\alpha\tau-2/T)},\quad
\label{c_2}
\end{eqnarray}
valid  for $|s_{\vec k}|^2\ll  n_{\vec k}\ll  1$.  Thus,  the original
effect  advocated  in  Refs.  \cite{Asakawa:1996xx,Asakawa:1998cx}  of
unlimitedly  large  correlations  with  increasing  $\Omega_{\vec{k}}$
resulting  in  the sudden  case  ($\tau=0$)  turns  into the  opposite
behavior,  once   the  duration   time  $\tau$  exceeds   the  inverse
temperature $1/T$.  Already  for charged pion pairs ($\pi^{+}\pi^{-}$)
with  $\Omega_{\vec  k}\approx  300$~MeV the  correlation  (\ref{c_2})
falls  below the $10^{-5}$  level for  transition times  $\tau$ beyond
3~fm\!/\!c, at typical  freeze-out temperatures of $T\approx 140$~MeV.
Since optical  potentials are  generally proportional to  the density,
the  characteristic  transition-time  scale  is rather  given  by  the
expansion-time scale \cite{Knoll:2008sc,Knoll:2009nf,CBM-Book:2010} in
nuclear collisions.  Its  full-width-half-maximum values generally lie
beyond 4 fm\!/\!c, leading  to even larger suppression factors.  Thus,
the pair creation process due to  the time variation of the mean field
becomes unmeasurably small in nuclear collisions.

The  discussed particle--antiparticle  correlation effect  rests  on a
coherent  single-particle picture.   Among others  it  largely ignores
collisional effects  that are known to dominate  the nuclear collision
dynamics.  Thus,  in individual events the one-body  field will depart
from  the  ensemble mean  due  to  fluctuations  caused by  stochastic
processes.  Such  microscopic processes can involve  much shorter time
scales  which,  however,  are  also  ultra violet  restricted  by  the
frequencies accessible in the  system.  For thermal systems this limit
is given by the  inverse temperature scale, i.e.  $\tau>1/T$, limiting
the  ``thermal''  production  of   {\em  hard}  probes,  such  as  the
particle--antiparticle pairs.  Such stochastic processes generally add
incoherently   to  the   two-particle  yield.    Therefore   they  can
appropriately  be  included  in  transport  codes  by  simulating  the
corresponding  microscopic collision  processes.   The resulting  pair
correlations, however,  are no  longer strictly back-to-back,  as they
are influenced by thermal motion.

The question, whether a process can  be treated in the sudden limit or
not, depends  on its intrinsic  quantum time scale resulting  from the
uncertainty principle  to $\tau_Q\approx1/\Delta E$,  where $\Delta E$
is the  energy transfer.  Driven by  a certain field,  the dynamics is
``sudden'',  if the  typical time  duration $\tau$,  during  which the
field changes, is short compared  to $\tau_Q$: then the wave functions
stay  ``inert''   across  the  transition   thereby  creating  several
components  in the  eigenstates of  the ``new''  Hamiltonian.   In the
opposite limit  $\tau\gg\tau_Q$ the  quantum states change  and adjust
adiabatically.   As  shown here  for  the  pair  creation case,  where
$\Delta  E=2\Omega_{\vec{k}}$,  the creation  of  states involving  an
energy  transfer  $\Delta  E\gg  1/\tau$  then  becomes  exponentially
suppressed.\\[-1cm]
\section*{Acknowledgement}
The   author    acknowledges   constructive   discussions    with   P.
Braun-Munzinger, B. Friman and D.N. Voskresensky.

\bibliography{References}
\end{document}